\documentclass[twocolumn,showpacs]{revtex4}

\usepackage{dcolumn}
\usepackage{amsmath}
\usepackage{graphicx}

\begin{document}

\newcommand \be {\begin{equation}}
\newcommand \ee {\end{equation}}
\newcommand \bea {\begin{eqnarray}}
\newcommand \eea {\end{eqnarray}}
\newcommand \nn {\nonumber}

\title{Crossover from entropic to thermal dynamics in glassy models}
\author{Eric M. Bertin}
\affiliation{CEA -- Service de Physique de l'\'Etat Condens\'e, Centre d'\'Etudes de Saclay, F-91191 Gif-sur-Yvette Cedex, France}

\date{\today}

\begin{abstract}
We study the emergence of a crossover from entropically driven to thermally activated dynamics in different versions of the `entropic' phase space model introduced by Barrat and M\'ezard, and previously considered in the zero temperature limit. We first focus on the low temperature ($T \leq T_g$) aging phase of the original model and show that a short time singularity appears in the correlation function for $T_g/2 < T < T_g$, leading to dynamical ultrametricity at $T=T_g$. We then consider the finite size version of this model, showing that the long time dynamics is always thermally activated beyond a size dependent crossover time scale. We also generalize the model, introducing a threshold energy so as to mimic a phase space composed of saddles above the threshold, and minima below. In this case, the crossover time scale becomes much smaller than the equilibration time, and both kinds of aging dynamics are successively found, inducing a non trivial aging scaling which does not reduce to the usual $t/t_w$ (or even $t/t_w^{\nu}$) one.

\end{abstract}


\maketitle

A major part of our understanding of the glass transition comes from Mode Coupling Theory (MCT) for structural glasses \cite{Gotze,Gotze-Houches} and from dynamical mean-field theory for spin glasses \cite{review,pspin}, which are the most sophisticated microscopic theories available in glassy physics. Apart from these theories, the energy landscape paradigm has raised a lot of interest in recent years \cite{Stil95,Heuer99,Deben01,Scior02}, due to the increasing power of computer simulations which now allow to implement the seminal ideas of Goldstein \cite{Goldstein}, specified later by Stillinger and Weber \cite{StiWeb}. These studies have led to associate the mode coupling transition temperature $T_c$, where MCT locates an ideal glass transition, with a threshold energy $E_{th}$ of geometric nature, separating regions of phase space dominated by saddle-points for $E>E_{th}$ from regions dominated by minima for $E<E_{th}$ \cite{Scior00,Broderix,Giardina,Keyes,Angelani03}. In this framework, the ideal glass transition found in MCT and in mean-field theories is explained on the one hand by the sharp separation between saddles and minima, and on the other hand by the divergence of the energy barriers surrounding minima \cite{Kurchan93}.

If one of these two conditions was not fulfilled, the ideal glass transition would be smeared out, which indeed happens in real glass formers \cite{Gotze-Houches}. In particular, if energy barriers remain finite, one may expect not only the behaviour of the relaxation time with temperature to be smoothed, but also correlation functions to reflect this dynamical crossover between a time regime when saddles are mainly explored and a later regime in which the system visits minima in majority. The key point here is that visiting minima corresponds to thermally activated dynamics where the system has to overcome energy barriers, which thus depends crucially on temperature, whereas visiting saddles is associated to a rather temperature independent kind of dynamics, sometimes called `entropic dynamics': the evolution of the system is mainly limited due to the lack of escape downwards directions, when visiting a saddle with only a few negative eigenvalues. Interestingly, this dynamical transition has already been observed in Lennard-Jones supercooled liquids \cite{Angelani,Reichman}, but a quantitative analysis of the associated crossover scales has not been achieved yet.

In this paper, we study quantitatively this crossover and its dependence upon relevant parameters in simple stochastic models of glassy dynamics. We first consider the phase space model introduced by Barrat and M\'ezard \cite{BM95} -- hereafter denoted as BM model -- which has known a recent surge of interest in the context of fluctuation dissipation relations \cite{Ritort,Sollich}, and study the aging regime of the correlation function $C(t_w,t_w+t)$ in the whole glassy phase $T<T_g$, generalizing previous results obtained in the limit $T \to 0$. We find a temperature independent long time behaviour ($t \gg t_w$), as well as a short time singularity ($t \ll t_w$) in the temperature range $T_g/2 < T < T_g$, which leads to dynamical ultrametricity \cite{CuKu} for $T=T_g$, as in the trap model \cite{DU01}.
We then study the correlation function in the {\it finite size} BM model, both in equilibrium and in the aging regime, which allows to evidence clearly a dynamical crossover from entropic to activated dynamics.
Finally, the BM model is generalized by introducing a threshold energy so as to mimic a phase space structure composed of saddles and minima. We show that the change of aging dynamics (from entropic to thermally activated) in this model leads to a non trivial scaling behaviour, and that the crossover time scale can be much smaller than the equilibration time.

\section{The Barrat-M\'ezard model}

One of the first attempts to describe aging dynamics with simple stochastic phase space models was the trap model proposed by Bouchaud \cite{Bou92}, with the aim to evidence a possible general scenario for glassiness, which emphasizes activation effects. A few years later, Barrat and M\'ezard \cite{BM95} proposed a somewhat similar model, but where thermal activation was replaced by an `entropic activation', i.e. by the fact that at low temperature less and less lower energy states could be reached when the system is already in a low energy state. In a continuous energy formulation, corresponding to an infinite number of states, the BM model is defined as a Markovian process where a `particle' (which usually corresponds to the representative point of a system in its phase space) is allowed to jump from one energy state to any other one. These energy states are distributed according to an a priori distribution $\rho(E)$, and transition rates $W(E'|E)$ from energy $E$ to energy $E'$ correspond to the Glauber choice:
\be
W(E'|E) = \frac{\Gamma_0\, \rho(E')}{1+e^{(E'-E)/T}}
\ee
where $\Gamma_0$ is a microscopic frequency scale, that we shall set to unity in the following.
The dynamics of the BM model is then described by the probability $P(E,t)$ to have energy $E$ at time $t$, which satisfies the following master equation:
\be \label{Master-Eq}
\frac{\partial P}{\partial t} = \int_{-\infty}^0 dE' \,[W(E|E') P(E',t)- W(E'|E) P(E,t)]
\ee
where for simplicity -- but without loss of generality since we are interested in the low energy dynamics -- we have restricted $\rho(E)$ to be non-zero only for $E<0$.

\subsection{Zero temperature results}

Contrary to the trap model, and to any model driven by thermal activation, the temperature can be set to zero without freezing completely the dynamics, leading to an evolution somewhat similar to a steepest descent dynamics in a high dimensional continuous energy landscape. A noticeable property of this model in the zero temperature limit, which has been studied in details in \cite{BM95}, is that the dynamics, expressed in terms of sojourn times instead of energies, becomes completely independent of the functional form of the distribution $\rho(E)$; in particular, $\rho(E)$ can be bounded or not. Defining the average sojourn time $\tau(E)$ at energy $E$ by the relation:
\be \label{def-tauE}
\frac{1}{\tau(E)} \equiv \int_{-\infty}^0 dE'\, W(E'|E)
\ee
which reduces for $T=0$ to $\tau(E)^{-1} = \int_{-\infty}^E dE'\, \rho(E')$, one finds that the a priori distribution $\psi_0(\tau)$ derived from $\rho(E)$ is given by $\psi_0(\tau) = \tau^{-2}\, \theta(\tau-1)$, independently of $\rho(E)$. As a result the dynamical distribution $p(\tau,t)$, which has been computed in \cite{BM95}, does not depend either on the shape of $\rho(E)$. It was found that $p(\tau,t)$ cannot reach a steady state, but exhibits for large $t$ a scaling form $p(\tau,t)=t^{-1} \phi_0(\tau/t)$, typical of the aging behaviour, with $\phi_0(u)$ given by:
\be \label{dist0T}
\phi_0(u) = \frac{1}{u^2}\, e^{-1/u}
\ee
Defining the correlation function $C(t_w,t_w+t)$ as the probability not to move between $t_w$ and $t_w+t$, which can be written:
\be \label{def-correl}
C(t_w,t_w+t) \equiv \int_{-\infty}^0 dE\, P(E,t_w)\, e^{-t/\tau(E)}
\ee
one finds for $T=0$ and large $t_w$:
\be
C(t_w,t_w+t) = \frac{t_w}{t_w+t}
\ee
Note that in the `short time' regime $1 \ll t \ll t_w$, the correlation function $C(t_w,t_w+t)$ behaves linearly, in the sense that $1-C \sim t/t_w$; this point will be important in the discussion of the finite temperature results.

\subsection{Aging at finite temperature}

If one chooses an exponential density of states $\rho(E)= \beta_g \exp(\beta_g E)\, \theta(-E)$, as is usually done in the trap model, then the Boltzmann equilibrium distribution becomes non normalizable for $T \le T_g \equiv 1/\beta_g$. In this case, the dynamical distribution drifts towards deeper and deeper energies, leading to the aging phenomenon, which can be evidenced for instance by the behaviour of the correlation function. In this section, we present the new results we have obtained in the aging regime for the finite temperature case, namely for $0 < T < T_g$. 

As was done for the trap model in \cite{Monthus}, one can then look for a scaling solution for the dynamical energy distribution. A natural scaling variable would be the ratio of the typical time $\tau(E)$ the system spends in a state with energy $E$ -- as defined in Eq.~(\ref{def-tauE}) -- to the time $t$ elapsed after the quench. Computing $\tau(E)$ for $0 < T < T_g$ and large $|E|$ yields:
\be \label{eq-tauE}
\tau(E) = \frac{\sin \pi \mu}{\pi \mu} \, e^{-E/T_g}
\ee
where $\mu \equiv T/T_g$.
We choose a scaling variable $u$ proportional to $\tau(E)/t$, defining it through $u^{-1}=t\, e^{E/T_g}$. This leads to a distribution $P(E,t)= \beta_g \, u\, \phi(u)$, where $\phi(u)$ is normalized by $\int_0^{\infty} du\, \phi(u) = 1$. Once these substitutions are made in Eq.~(\ref{Master-Eq}), the following equation is obtained:
\be \label{eqn-scal}
u^2 \phi'(u) + (u-J)\, \phi(u) = -\frac{1}{u} \int_0^{\infty} dv \, \frac{\phi(v)}{1+(v/u)^{1/\mu}}
\ee
where $J \equiv \pi \mu / \sin \pi \mu$.
The asymptotic behaviour of the distribution $P(E,t)$ can be extracted rather easily from this equation. Let us analyse for instance the large $u$ behaviour of $\phi(u)$ -- or large $|E|$ behaviour of $P(E,t)$. In this case, the integral in the r.h.s of Eq.~(\ref{eqn-scal}) reduces to:
\be \label{integ-phi}
\int_0^{\infty} dv\, \frac{\phi(v)}{1+(v/u)^{1/\mu}} \simeq \int_0^{\infty} dv\, \phi(v) = 1 \qquad u \gg 1
\ee
so that Eq.~(\ref{eqn-scal}) reads, neglecting also $J$ with respect to $u$ and calling $\phi_l(u)$ the asymptotic large $u$ expression of $\phi(u)$:
\be
u^2 \phi_l'(u) + u\, \phi_l(u) = -\frac{1}{u}
\ee

\begin{figure}[t]
\centering\includegraphics[width=8.5cm]{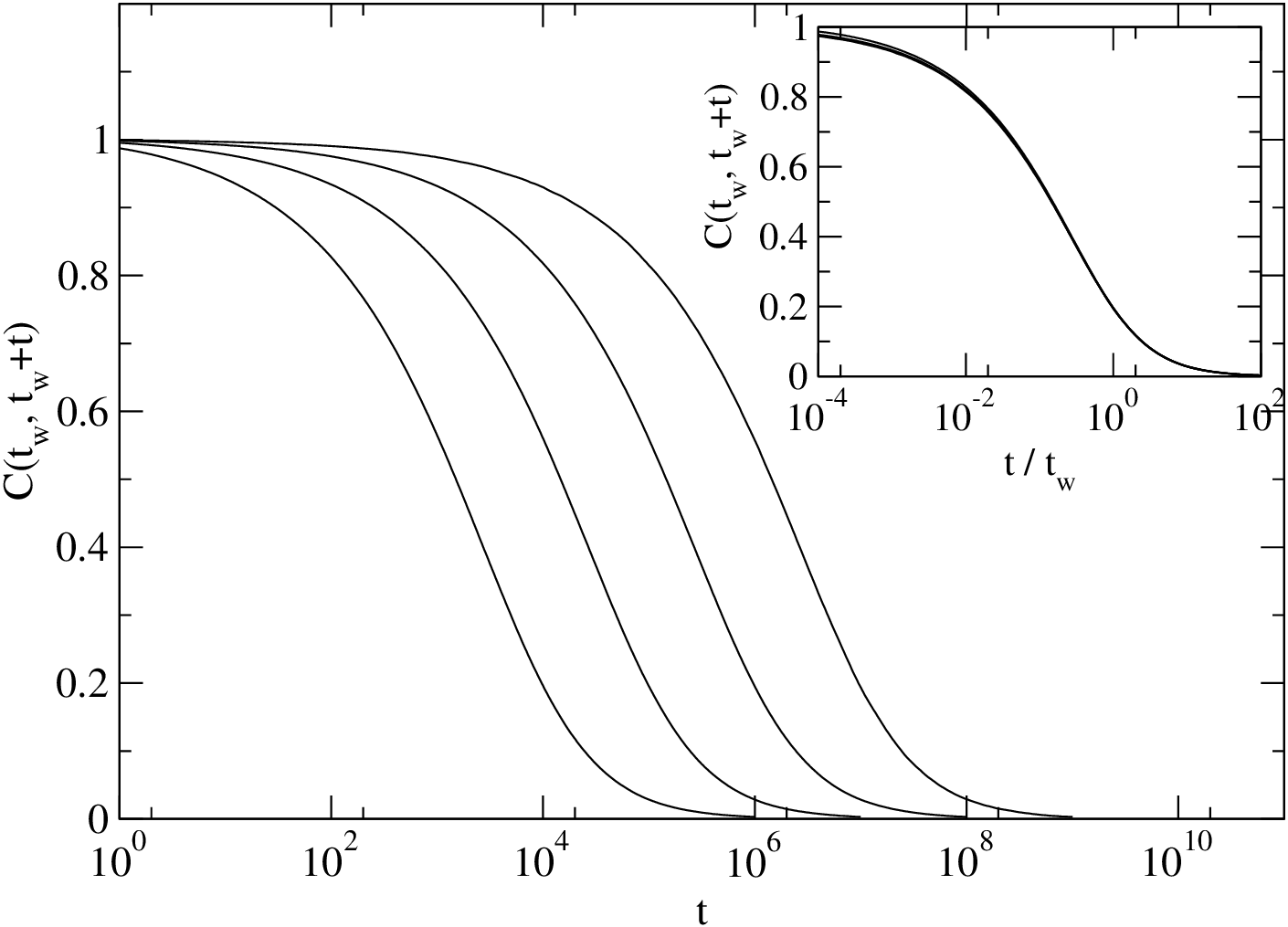}
\caption{\sl Plot of the aging correlation function $C(t_w,t_w+t)$ of the BM model, for $t_w=10^4$, $10^5$, $10^6$ and $10^7$ (from left to right), and temperature $\mu=0.7$. Inset: correlation plotted as a function of the rescaled time $t/t_w$, showing a very good collapse of the data, apart from small deviations for $t \ll t_w$ due to finite time effects.
}
\label{fig-ag-scal}
\end{figure}

\noindent
The general solution of this equation reads:
\be
\phi_l(u) = \frac{A}{u} + \frac{1}{u^2}
\ee
The constraint that $\phi(u)$ must be normalizable requires that $A=0$, so that $\phi_l(u)=u^{-2}$. Interestingly, we find here the \textit{exact} asymptotic expression for $\phi(u)$, including the prefactor -- equal to $1$ -- although we performed only an asymptotic analysis on a linear equation. Note also that taking into account the leading correction to the approximation made in Eq.~(\ref{integ-phi}) leads to $\phi_l(u) - u^{-2} \sim u^{-3}$.
Neglecting these corrections, one thus finds that $P(E,t)$ is equal to $\beta_g \, t \, e^{E/T_g}$ -- i.e. proportional to $\rho(E)$ -- for $e^{|E|/T_g} \gg t$.

In the opposite case $u \ll 1$, the equation governing the asymptotic expression $\phi_s(u)$ of $\phi(u)$ reads:
\be
u^2 \phi_s'(u) - J \phi_s(u) = -\int_0^{\infty} dz \, \frac{\phi_s(uz)}{1+z^{1/\mu}}
\ee
Searching for a power law solution $\phi_s(u)=a\, u^{\alpha}$, the above equation is satisfied only if:
\be \label{eq-alpha}
J = \int_0^{\infty} dz\, \frac{z^{\alpha}}{1+z^{1/\mu}}
\ee
since the first term $u^2 \phi_s'(u)$ can be neglected in this case. Replacing by the respective values of both sides, one has:
\be \label{eqn-sol}
\frac{\pi \mu}{\sin \pi \mu} = \frac{\pi \mu}{\sin (1+\alpha) \pi \mu}
\ee
The convergence of the integral appearing in Eq.~(\ref{eq-alpha}) requires that $0<(1+\alpha)\mu<1$. In this range, Eq.~(\ref{eqn-sol}) admits two solutions: $\alpha=0$ and $\alpha=1/\mu-2$. Since the zero temperature limit of the model is regular, the finite temperature solution must converge towards the zero temperature distribution given by Eq.~(\ref{dist0T}), which decays faster than any power law for $u \to 0$, so that one expects $\alpha \to \infty$ for $\mu \to 0$. The correct solution for $\phi(u)$ is thus:
\be \label{phi-short}
\phi(u) \sim u^{1/\mu-2} \qquad u \ll 1
\ee
So the energy distribution behaves at small $|E|$ -- i.e. for $e^{|E|/T_g} \ll t$ -- as:
\be
P(E,t) \sim t^{1-1/\mu}\, e^{(1-1/\mu)E/T_g} \sim t^{1-1/\mu}\, \rho(E)\, e^{-E/T}
\ee
which means that small energies are almost equilibrated.

Turning to the correlation function $C(t_w,t_w+t)$ defined above, Eq.~(\ref{def-correl}) can be rewritten in terms of the scaling function $\phi(u)$, using also Eq.~(\ref{eq-tauE}):
\be
C(t_w,t_w+t) = \int_0^{\infty} dv\, \phi(v) e^{-Jt/vt_w}
\ee
which shows that the correlation function exhibits a simple aging form:
\be \label{eq-C-aging}
C(t_w,t_w+t) = \mathcal{C}_{\textrm{ag}}(t/t_w)
\ee
Using the above asymptotic results for $\phi(u)$, one can find the short time and long time behaviour of the correlation function. In particular, for times $t \gg t_w$, $C(t_w,t_w+t)$ is given by:
\be \label{eq-cor-late}
C(t_w+t,t_w) \simeq \frac{\sin \pi \mu}{\pi \mu}\, \frac{t_w}{t} \qquad t \gg t_w
\ee
for any $\mu<1$, so that the `tail' of the correlation function appears to be a temperature independent power law, apart from the prefactor. This point will be important for later discussions.

The short time behaviour is computed from:
\be
1-C(t_w,t_w+t) = \int_0^{\infty} dv\, \phi(v)\, (1-e^{-Jt/vt_w})
\ee
For $t \ll t_w$, one can try to linearize the exponential in the last factor. Using Eq.~(\ref{phi-short}), one sees that the resulting integral converges at its lower bound only if $\mu < \frac{1}{2}$; in this case, the rescaled correlation function $\mathcal{C}_{\textrm{ag}}(u)$ is then regular when $u \to 0$ (i.e. $1-\mathcal{C}_{\textrm{ag}}(u) \sim u$), just as in the zero temperature case. On the contrary, for $\frac{1}{2} < \mu < 1$, a singularity appears, and one then finds:
\be
1-\mathcal{C}_{\textrm{ag}} \left( \frac{t}{t_w} \right) \sim \left(\frac{t}{t_w} \right)^{(1-\mu)/\mu} \qquad t \ll t_w
\ee
This property can be interpreted as follows: the average energy $\langle E' \rangle_{\textsc{e}}$ reached in a transition from $E$ to $E'$ reads:
\be
\langle E' \rangle_{\textsc{e}} \equiv \int_{-\infty}^0 dE'\, E'\, W(E'|E) = E-\frac{\pi \mu}{\tan \pi \mu} \, T_g
\ee
For $\mu < \frac{1}{2}$, $\langle E' \rangle_{\textsc{e}} < E$ so that the dynamics resembles the zero temperature one: at each step, the energy is lowered on average. On the contrary, for $\mu > \frac{1}{2}$, $\langle E' \rangle_{\textsc{e}} > E$ and the energy is {\it raised} on average at each transition. So the slow drift of the average energy $\overline{E}(t)$ as a function of time results in this case from a kinetic effect: the particle stays a longer time on lower energy states. Actually, the particle typically jumps a lot of times among high energy states before reaching a low energy one, then jumps back to high energies and so on. This scenario is reminiscent of what happens is the trap model, for which $\langle E' \rangle_{\textsc{e}} = T_g$ independently of the starting energy $E$, and is presumably responsible for the onset of the short time singularity. Note however that in the BM model, \textit{both} entropic and thermal dynamics come into play for $1/2<\mu<1$, the latter being rather irrelevant for $\mu<1/2$, but thermal activation remains in any case an auxiliary mechanism, the main one being entropic. Indeed, this can be seen from the typical sojourn time $\tau(E)$ -- see Eq.~(\ref{eq-tauE}) -- which is always proportional to $e^{-E/T_g}$, and not $e^{-E/T}$ as would be the case for a purely thermal dynamics. So the present scenario is a bit different from the usual landscape picture, since the entropic channel exists here even at very low energies. We shall see in the next sections how to generalize the model in order to account for a complete vanishing of downwards escape directions.

\begin{figure}[t]
\includegraphics[width=8.5cm]{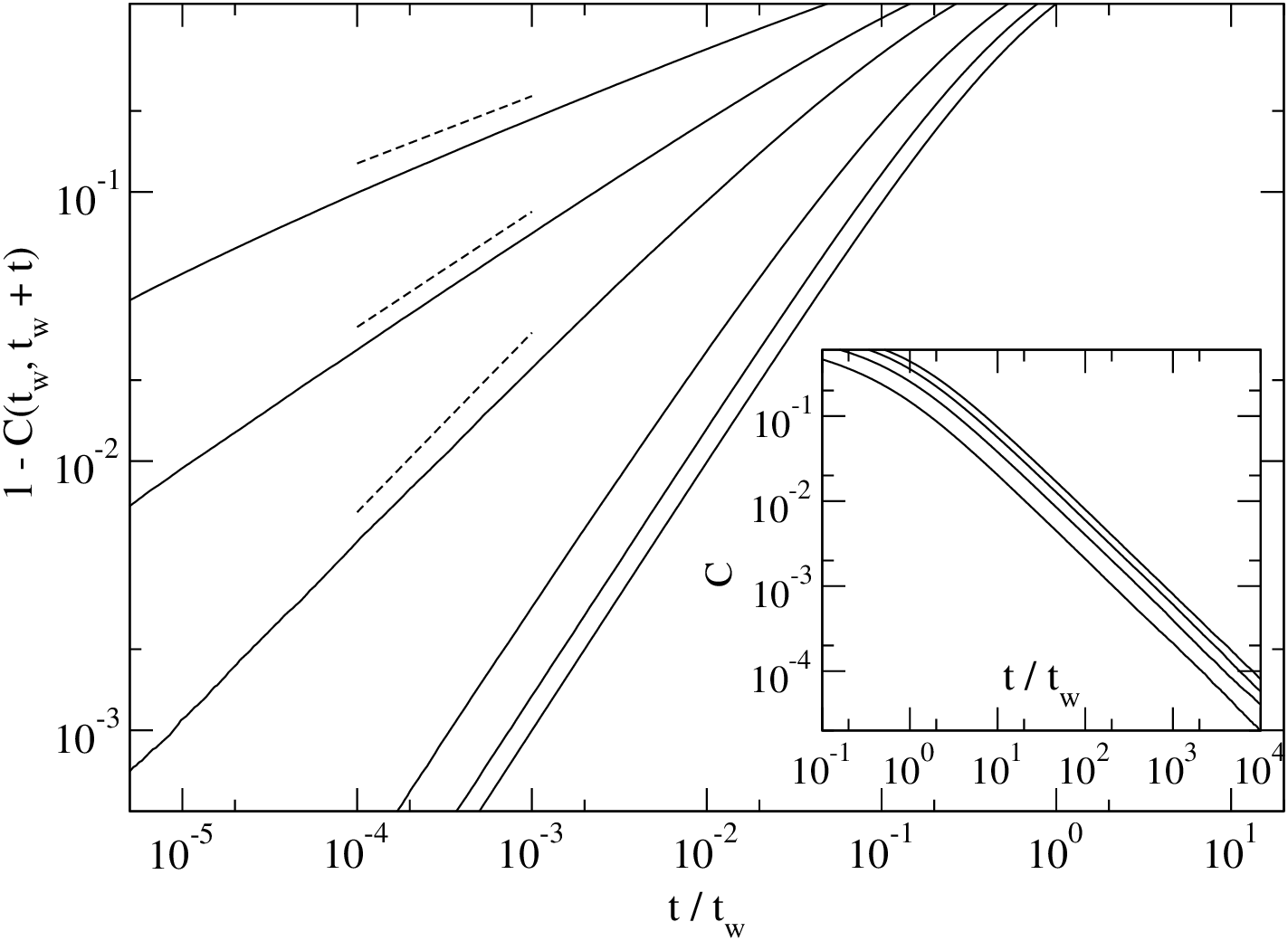}
\caption{\sl Short time behaviour of the correlation function; $1-C(t_w,t_w+t)$ is plotted as a function of $t/t_w$ for several temperatures $\mu=0$, $0.2$, $0.4$, $0.6$, $0.7$ and $0.8$ (from right to left), showing the onset of a short time singularity for $\frac{1}{2} < \mu < 1$ with an exponent $(1-\mu)/\mu$ (dashed lines). For $\mu < \frac{1}{2}$, $1-C$ is linear for $t \ll t_w$. Inset: long time tail of the correlation for $\mu=0.2$, $0.4$, $0.6$ and $0.8$ (from right to left), showing a power law decay $C \sim (t/t_w)^{-1}$.
}
\label{fig-ag-asymp}
\end{figure}

We now present numerical results in order to confirm the above analysis. For computational reasons, simulations were performed with a slightly different version of the model, in which the transition rates are given by:
\bea
\tilde{W}(E'|E) &=& \Gamma_0 \, \rho(E') \qquad \qquad \qquad \quad \, E' \leq E\\
\tilde{W}(E'|E) &=& \Gamma_0 \, \rho(E')\, e^{-(E'-E)/T} \qquad E' > E
\eea
These transition rates $\tilde{W}(E'|E)$ behave qualitatively as the usual Glauber rates, so that one expects this difference not to affect the scaling and power law behaviours, but only the numerical prefactors. One needs first to check the aging form -- Eq.~(\ref{eq-C-aging}) -- of the correlation function.
This is done on Fig.~\ref{fig-ag-scal}, which displays the correlation function $C(t_w,t_w+t)$ for $\mu=0.7$ and several waiting times, $t_w=10^4$, $10^5$, $10^6$ and $10^7$. The rescaling, shown in inset, is very good, with only small deviations in the short time regime $t \ll t_w$ where finite time effects become stronger. We also test the predicted exponents for the short time and late time behaviour of the correlation, as shown on Fig.~\ref{fig-ag-asymp}. One clearly sees the linear behaviour of $1-C$ for $\mu < \frac{1}{2}$, as well as the predicted singularity with exponent $(1-\mu)/\mu$ for $\frac{1}{2} < \mu < 1$. It is however difficult to find a genuine straight line for this short time singularity since $t$ must satisfy $1 \ll t \ll t_w$, and such a separation of scales is hard to obtain numerically (note that $t_w$ is already as large as $10^7$). The late time regime is easier to evidence quantitatively, since the only condition is $t \gg t_w$; the corresponding power law behaviour proportional to $(t/t_w)^{-1}$, with a temperature independent exponent, is clearly evidenced on the inset of Fig.~\ref{fig-ag-asymp}. As explained above, we cannot test quantitatively the prefactor predicted in Eq.~(\ref{eq-cor-late}) since the numerics deals with a slightly different model, but we can see that the prefactor found in simulations decreases with temperature, in qualitative agreement with the prediction.

\subsection{Dynamical ultrametricity for $T=T_g$}

The study of the behaviour of the model right at the glass transition temperature $T=T_g$ appears to be easier to solve, although the scaling form is a bit more subtle. In order to simplify the notations, we choose in this section $T_g$ as the energy (and temperature) unit. It can be shown that the correct scaling variable $u$ is given in this case by $u^{-1}=e^E\,t \ln t$, and that the associated scaling function $\varphi(u)$ must be related to $P(E,t)$ through:
\be
P(E,t) \simeq \frac{1}{\ln t}\, u\, \varphi(u)
\ee
The normalization condition $\int_{-\infty}^0 dE\, P(E,t)$ translates into:
\be \label{eqn-norm}
\frac{1}{\ln t} \int_{(t \ln t)^{-1}}^{\infty} du\, \varphi(u) \to 1 \qquad t \to \infty
\ee
which requires that $\varphi(u) \simeq u^{-1}$ for $u \to 0$. The master equation then becomes in the limit of large $t$:
\be \label{eqn-scal-crit}
u^2 \varphi'(u) + (u-1)\, \varphi(u) = -\frac{1}{u}
\ee
The condition that $\varphi(u)$ must be normalizable in the sense of Eq.~(\ref{eqn-norm}) selects a unique solution, which reads:
\be
\varphi(u)=\frac{1}{u}(1-e^{-1/u})
\ee
so that $P(E,t)$ is given by
\be
P(E,t) = \frac{1}{\ln t} \left[ 1-\exp(-e^E t \ln t) \right]
\ee
From this expression, one can compute the correlation $C(t_w,t_w+t)$ at the critical temperature:
\be
C(t_w,t_w+t) = \int_{-\infty}^0 dE\, P(E,t_w)\, e^{-t/\tau(E)}
\ee
with $\tau(E)$ given by:
\be
\frac{1}{\tau(E)} \equiv \int_{-\infty}^0 \frac{e^{E'} \, dE'}{1+e^{E'-E}} = e^E \ln (1+e^{-E})
\ee
which reduces to $|E|\, e^E$ for large $|E|$ ($E<0$). So the correlation reads, for large $t_w$ and $t$:
\bea
C(t_w,t_w+t) &=& \frac{1}{\ln t_w} \int_{-\infty}^0 dE\, \exp(E\, t\, e^E)\\ \nonumber
&& \qquad \times \left[ 1-\exp(-e^E t_w \ln t_w) \right]
\eea
From this expression, one can deduce that the relevant scaling variable is no longer $t/t_w$, but rather $\omega = \ln t/\ln t_w$, in the sense that when taking the infinite $t_w$ limit keeping $\omega$ fixed, $C(t_w,t_w+t)$ converges to a function $\tilde{\mathcal{C}}(\omega)$ given by:
\be
\tilde{\mathcal{C}}(\omega)=1-\omega \quad (0 \leq \omega <1), \quad \tilde{\mathcal{C}}(\omega)=0 \quad (\omega \geq 1)
\ee
This is precisely the asymptotic form found in the trap model at the glass transition temperature, which was shown to satisfy dynamical ultrametricity \cite{DU01,CuKu}. So we see that dynamical ultrametricity is indeed independent of the type of dynamics considered -- entropic or thermal activation -- but is rather governed by the exact balance of the Boltzmann weight and of the exponential density of states which is valid only right at the glass transition temperature. Interestingly, the differential equation -- Eq.~(\ref{eqn-scal-crit}) -- satisfied by the scaling function $\varphi(u)$ appears to be the same in the BM model as in the trap model -- although another approach was used in \cite{DU01} -- leading to the same scaling functions $\varphi(.)$, even though they are derived from two different master equations. Note however that the scaling variables are slightly different, $u=(t \ln t \, e^E)^{-1}$ in the BM model, and $v=(t\, e^E)^{-1}$ in the trap model.

\section{Finite size equilibrium}

In realistic models of glasses, the structure of the global phase space is expected to be highly non trivial, since it has to encode the finite dimensionality of the physical space. So it appears tempting, at least as a first approximation, to think of a macroscopic system as a collection of small independent subsystems, the size of which would be of the order of the coherence length \cite{JP-Houches,Biroli} -- see also \cite{Vincent} for an experimental investigation of this point. This scenario has indeed been supported by recent simulations \cite{Heuer02} on Lennard-Jones supercooled liquids, showing that a system with $130$ particles has to a good approximation the same dynamical behaviour as two non interacting systems composed of $65$ particles each \endnote{Note that this decomposition into a large number of `elementary correlated cells' should not be confused with the usual procedure invoked to justify self-averaging properties in disordered systems, where each cell is yet assumed to be very large.}.

In this context, finite size versions of the usual -- trap and BM -- phase space models appear to be particularly relevant, and the simplest case to study is the equilibrium one. In order to make contact with more realistic models, let us emphasize that `states' in the present case should be associated with singular points in phase space, namely saddles or minima, so that one expects their number $M$ to grow exponentially with the number $N$ of physical degrees of freedom:
\be \label{eq-MvsN}
M \sim M_0\, e^{\alpha N}
\ee
where $\alpha$ is a numerical constant. A lot of studies have investigated the value of $\alpha$ as a function of temperature when considering only minima and have found $\alpha \ll 1$ \cite{BrayMoore,Rieger,Heuer99}. Note however that, bearing in mind the above picture, $N$ should be itself a function of temperature since it depends on the coherence length $\xi(T)$ through $N \sim \xi(T)^d$.

Considering $M$ energy states randomly chosen from an exponential distribution $\rho(E)$, the equilibrium distribution concentrates onto the lowest energy states, so that the correlation function decays on a time scale corresponding to the typical largest sojourn time $\tau_{\rm max}$: $C_{eq}(t,M) = \mathcal{ C}(t/\tau_{\rm max})$. The value of $\tau_{\rm max}$ can be estimated using a scaling argument: in a finite size model, one can imagine that all the sojourn times are chosen at random according to the probability distribution $\psi(\tau)$ associated to the \textit{infinite} size model. In the exponential trap model, $\psi_{\textsc{tm}}(\tau) = \mu/\tau^{1+\mu}$ and this picture is indeed exact, since the relation $\tau = e^{E/T}$ holds locally, at the trap level: one has then $\tau_{\rm max} \sim M^{1/\mu}$. On the contrary, in the BM model, the relation given in Eq.~(\ref{eq-tauE}) is valid only in the infinite size limit, and should be only an approximation in the finite $M$ case. Still, one can hope that this approximation is able to give the correct scaling of $\tau_{\rm max}$ with $M$: $\psi(\tau) \sim 1/\tau^2$ then yields $\tau_{\rm max} \sim M$.

An interesting question naturally arises about the finite size equilibrium dynamics in the low temperature phase. Given that for $T \leq T_g$ the continuous energy equilibrium distribution $P_{eq}(E) \propto \rho(E)\, e^{-E/T}$ becomes non normalizable, a cut-off scale necessarily appears in order to regularize the distribution $P_{eq}(E)$. One way to introduce such a cutoff is precisely to consider a finite number of states. Otherwise, the spontaneous cutoff is generated by the dynamics itself, since it is given by the time $t_w$ elapsed after the quench. So one can wonder whether or not the dynamics is qualitatively the same whatever the nature of the cut-off; we shall see that the answer depends on the model considered.

In the exponential trap model in the low temperature phase $\mu < 1$, with a finite number $M$ of traps, the correlation $C_{eq}^{\textsc{tm}}(t,M)$ is thus a function $\mathcal{C}_{\textsc{tm}}(t/M^{1/\mu})$ of the rescaled time $t/M^{1/\mu}$, which can be computed in the asymptotic regimes $t \ll M^{1/\mu}$ and $t \gg M^{1/\mu}$, leading to the interesting result that the behaviour is similar to that found in the aging case, replacing the cut-off $t_w$ by $M^{1/\mu}$. However, the prefactors are different from that found in the aging case \cite{Dean}. Namely, one finds:
\bea
1-C_{eq}^{\textsc{tm}}(t,M) &\simeq& \kappa_s \, \left( \frac{t}{M^{1/\mu}} \right)^{1-\mu} \quad t \ll M^{1/\mu}\\
C_{eq}^{\textsc{tm}}(t,M) &\simeq& \kappa_l \,\left( \frac{t}{M^{1/\mu}} \right)^{-\mu} \quad \; \, t \gg M^{1/\mu}
\eea
where the prefactors $\kappa_s$ and $\kappa_l$ are given by:
\bea
\kappa_s &=& \frac{1}{1-\mu} \, \Gamma(\mu) \, \Gamma(1/\mu) \, \Gamma(1-\mu)^{-1/\mu} \\
\kappa_l &=& \sqrt{2}^{1-2\mu} \,\mu \, \sqrt{\pi} \; \Gamma \left(\frac{1}{2}+\mu \right)
\eea
This shows that in the trap model, the asymptotic behaviour of the rescaled correlation function for $\mu < 1$ is the same for $t$ much smaller or much greater that the cut-off time scale, whatever the origin of this scale, namely static (number of sites) or dynamic (waiting time). It would be interesting to see if this property also holds for the BM model.

\begin{figure}[t]
\centering\includegraphics[width=8.5cm]{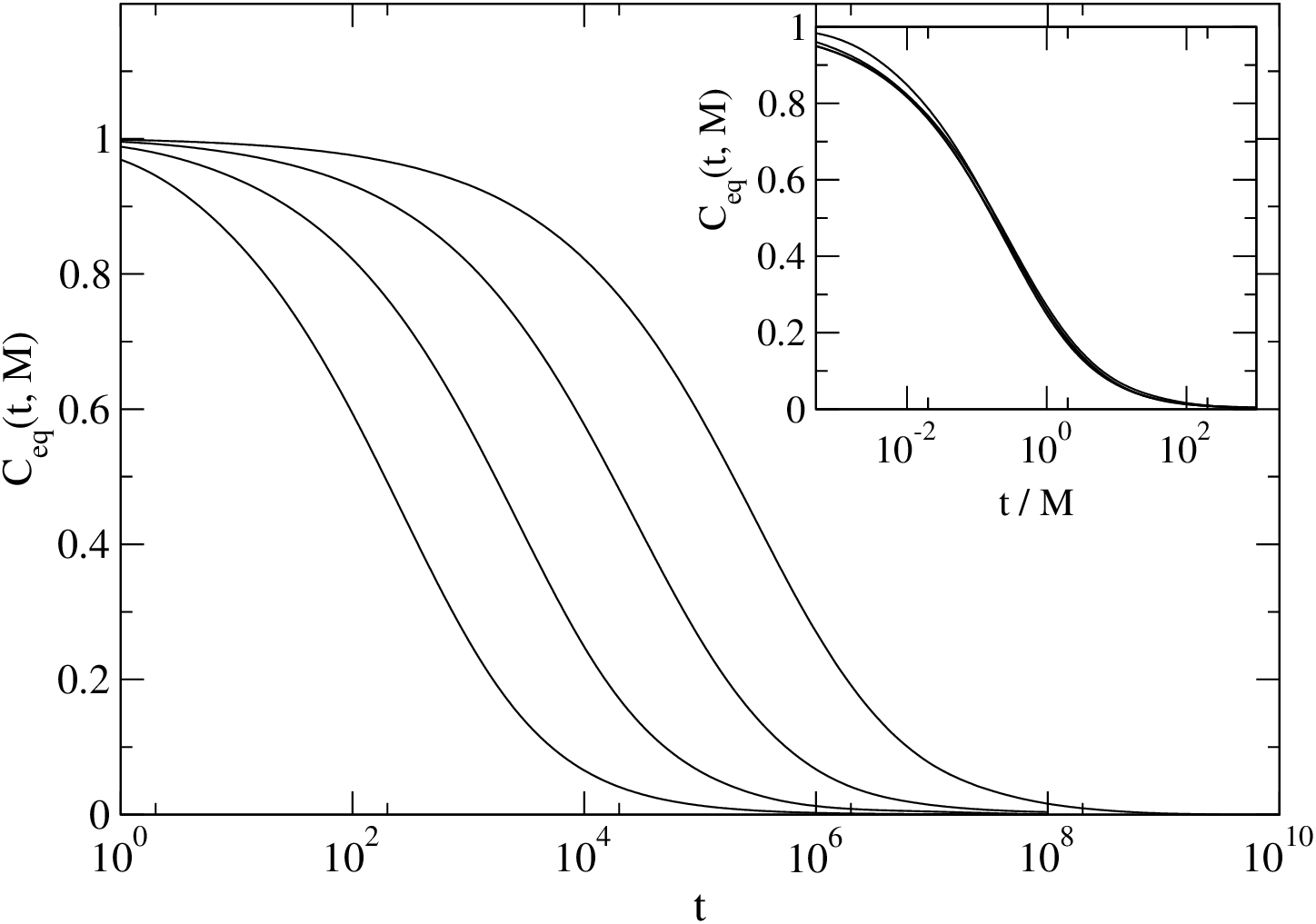}
\caption{\sl Plot of the finite size equilibrium correlation function $C_{eq}(t,M)$ of the BM model, for $M=10^3$, $10^4$, $10^5$ and $10^6$, and temperature $\mu=0.7$. Inset: correlation plotted as a function of the 1 time $t/M$, showing a good collapse of the data, except for short times where the $M=10^3$ curve does not rescale completely.
}
\label{fig-eq-scal}
\end{figure}

To this aim, we first check numerically the proposed scaling in $t/M$ for the BM model, using in this case the natural version of the model, with Glauber rates. Fig.~\ref{fig-eq-scal} displays $C_{eq}(t,M)$ for several values of $M$, namely $M=10^3$, $10^4$, $10^5$ and $10^6$; the inset shows that the data collapse well when plotted as a function of $t/M$, with again small deviations in the short time regime which requires larger system sizes to reach the asymptotic scaling form.

\begin{figure}[t]
\centering\includegraphics[width=8.5cm]{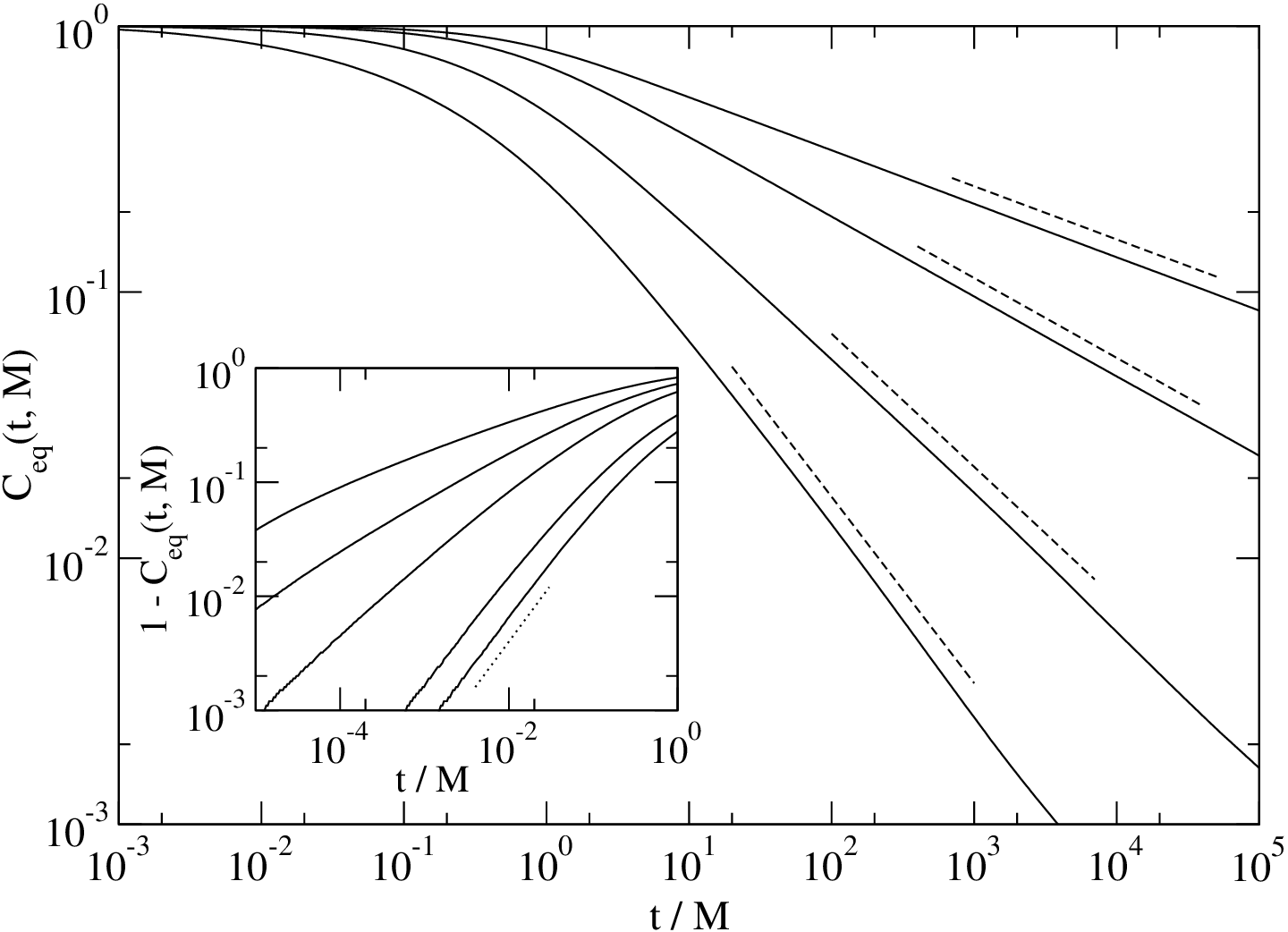}
\caption{\sl Asymptotic large time behaviour ($t \gg M$) of the correlation $C_{eq}(t,M)$ with $M=10^3$ for temperatures $\mu=0.2$, $0.3$, $0.5$ and $0.7$ (from top to bottom), showing a temperature dependent power law tail $C_{eq}(t,M) \sim (t/M)^{-\mu}$ (dashed lines), typical of activated dynamics. Inset: short time behaviour ($t \ll M$) of the correlation, with $M=10^6$, $\mu=0.3$, $0.4$, $0.6$, $0.7$ and $0.8$, showing the onset of the same short time singularity as in the aging regime (see Fig.~\ref{fig-ag-asymp}; dotted line: slope $-1$).
}
\label{fig-eq-asymp}
\end{figure}

Turning to the asymptotic behaviour of the correlation, one sees on Fig.~\ref{fig-eq-asymp} that the late time behaviour of $C_{eq}(t,M)$ is no longer temperature independent, as in the aging case, but depends strongly on temperature, and the slope is precisely equal, to numerical precision, to $-\mu$. This is a strong evidence for a thermally driven dynamics at large times, which corresponds to the fact that the particle can only escape from the deepest traps through thermal activation. On the contrary, the short time singularity is not modified with respect to the aging case, as can be seen in the inset of Fig.~\ref{fig-eq-asymp}. This was expected as the short time dynamics is dominated by `intermediate' energy states -- i.e. the highest states among the low energy ones onto which the equilibrium measure concentrates -- from which one can escape downwards.

As a result, a crossover appears in the equilibrium {\it finite size} correlation 0 between an entropic behaviour at short times and an activated one at long times. A natural question at this stage would be to see if this crossover is a peculiarity of the equilibrium case, or if it can be found also in dynamical regimes.

\section{Crossover in the aging regime}

Thermally activated phenomena are known to play an important r\^ole in glassy physics, and numerous experimental studies have focused on the effect of temperature on glasses (see e.g.~\cite{Tarjus} for structural glasses and \cite{PRB02} for spin glasses). Theoretical works also support this view, since the ideal glass transition predicted by MCT has been shown to be smeared out in real glasses by activated processes \cite{Gotze,Gotze-Houches}. However, in structural glasses, these activated processes are usually assumed to be relevant only on time scales close to the equilibration time $\tau(T)$ -- which becomes very large below the glass transition temperature -- and not in the aging regime which corresponds to shorter time scales. Indeed, recent simulations on Lennard-Jones liquids seem to validate a mean-field-like scenario with a well-defined effective temperature \cite{JLBarrat}, at odds with the behaviour of thermally driven models like the trap model \cite{Fielding} -- except in some very special cases \cite{Ritort}.

One might thus think that the crossover from entropic to thermal dynamics observed in numerical simulations \cite{Angelani,Reichman} is only due to the small size of the systems considered, and that the associated time scale should be very large for macroscopic systems, as shown in the mathematical analysis of the dynamics of the random energy model \cite{BenArous,Derrida}. Still, the arguments presented above suggest that for a finite dimensional system, this time scale should be governed by the coherence length rather than the global size of the system. One can also wonder whether the crossover time scale could be possibly much smaller than the equilibration time under some conditions.

\begin{figure}[t]
\centering\includegraphics[width=8.5cm]{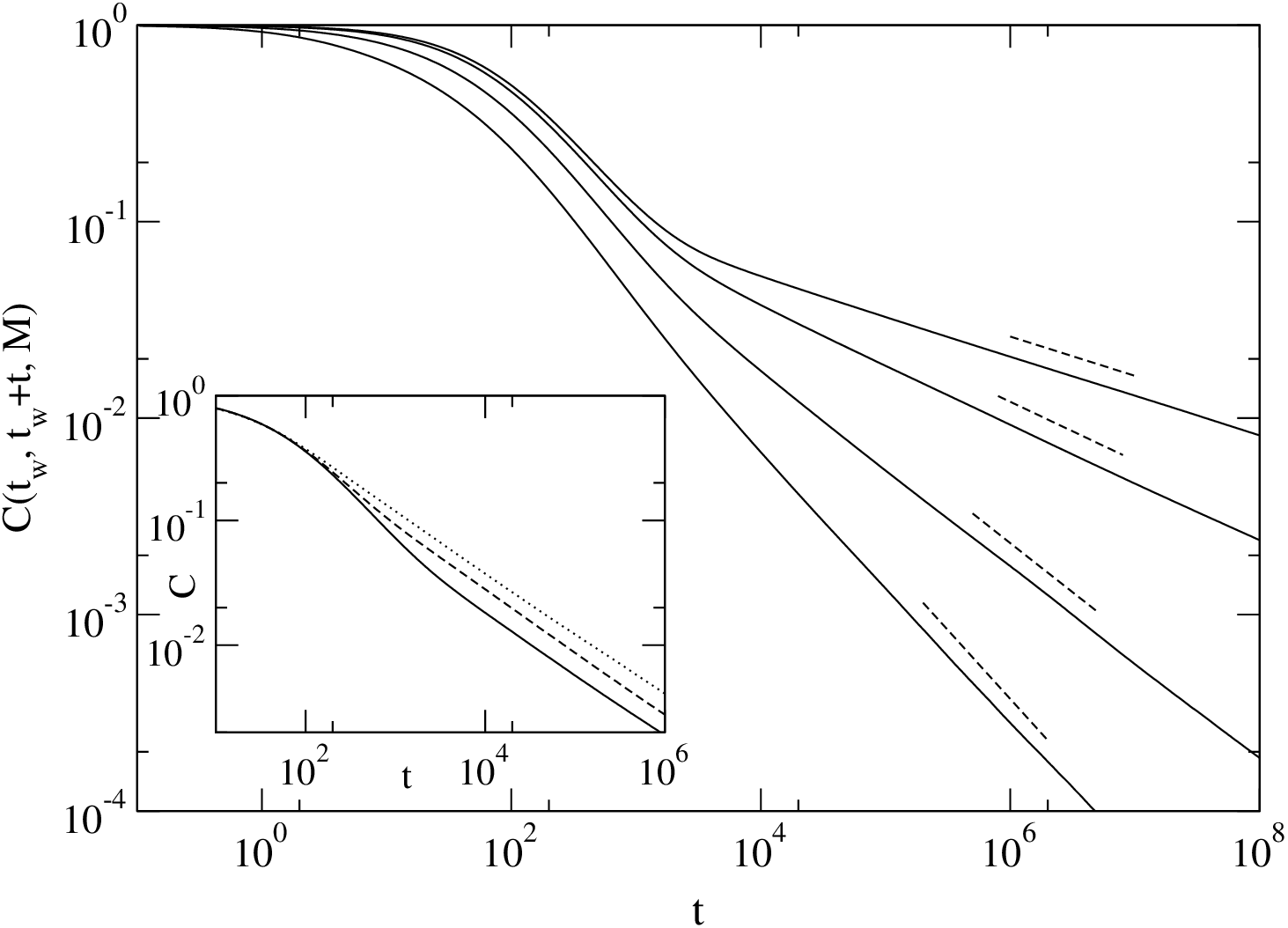}
\caption{\sl Correlation $C(t_w,t_w+t,M)$ in the out of equilibrium regime of the finite size model, for $M=10^3$, $t_w=10^2$ and $\mu=0.2$, $0.3$, $0.5$ and $0.7$. One sees a crossover from a rather temperature independent (entropic) regime to a temperature dependent (activated) one with exponent $-\mu$ (dashed lines). Inset: Effect of a finite connectivity Z, for $M=10^3$, $t_w=10^2$ and $\mu=0.5$; dotted line: $Z=100$, dashed line: $Z=300$ and full line: fully connected case. Reducing the connectivity decreases the crossover time scale.
}
\label{fig-M-aging}
\end{figure}

In this section, we study the onset of a dynamical crossover from entropic to activated behaviour in the aging regime of the correlation function. To this aim, we consider the finite size BM model as well as a generalization including a threshold energy below which all states are isolated minima, and show that the dynamical crossover time is equal to the equilibration time only for the fully connected model, and becomes much smaller as soon as a non trivial phase space structure is taken into account.

\subsection{Aging in a finite size system}

A natural generalization of the analysis performed in the previous section is to study the aging regime $1 \ll t_w \ll M$ of the finite size BM model, with the aim to evidence a crossover in the correlation $C(t_w,t_w+t,M)$ from the usual BM aging form for $t \ll M$ to an activated behaviour
for $t \gg M$. So the correlation function is now characterized by a couple of time scales, $t_w$ and $M$, instead of only one of these.

Numerical results for the correlation function $C(t_w,t_w+t,M)$ are shown on Fig.~\ref{fig-M-aging} for $M=10^3$, $t_w=10^2$ and various temperatures ($\mu=0.2$, $0.3$, $0.5$ and $0.7$). Apart from a global shift, the `early parts' of the curves (i.e. times $t$ such that $t_w \lesssim t \lesssim M$) are almost temperature independent, which is characteristic of an entropic kind of dynamics, as we have seen in the previous sections. On the contrary, the late ($t \gtrsim M$) power law decay, with an exponent $-\mu$ (dashed lines) is the fingerprint of thermally activated dynamics.

As we have only considered up to now fully connected models, one can wonder whether a finite connectivity $Z \ll M$ would strongly or not modify the previous results. Although a detailed study of the BM model with finite connectivity is beyond the scope of the present work, we comment here on the results of preliminary simulations on random graphs with finite average connectivity. In particular, decreasing the connectivity does not affect much the overall decay time of the correlation function, in equilibrium as well as in the aging regime. Still, out-of-equilibrium numerical data presented on the inset of Fig.~\ref{fig-M-aging} for $M=10^3$ and different values of $Z$, indicate that reducing the connectivity tends to decrease the crossover time scale beyond which thermal activation becomes dominant. This could have been expected, since a finite connectivity generates a lot of local minima, from which one can escape only through thermal activation. So we can guess that in a realistic glass former, even though the `elementary correlated cell' can contain a lot of degrees of freedom, the effect of connectivity can further reduce the crossover time scale, suggesting that activated phenomena should be relevant on accessible time scales.

Although a quantitative study of the finite size BM model in the out-of-equilibrium regime would be of great interest, it would require a larger scale separation than the one we obtained, which is hard to reach for computational reasons. We thus propose another version of the BM model, in which one can safely take the limit $M \to \infty$ without eliminating the dynamical crossover we are presently interested in. This can be done by taking into account a non trivial structure of phase space.

\subsection{Effect of a threshold energy}

Coming back to the paradigm of the energy landscape with a sharply defined separation in energy between saddles and minima \cite{Scior00,Broderix,Giardina,Keyes,Angelani03}, one could try to mimic this phase space structure by introducing by hand a threshold energy $E_{th}$ acting as a kinetic constraint, such that direct transitions between states with energies $E$ and $E'$ are forbidden if $E$ and $E'$ are both below $E_{th}$, and otherwise unchanged with respect to the usual BM model. In other words, states with energy $E>E_{th}$ can be considered as saddles, from which all other states are accessible, whereas states with energy $E<E_{th}$ represent isolated minima from which the system can escape only by jumping to a saddle, before eventually reaching another (possibly deeper) minimum. Such a definition could be applied to a finite size model, and the equilibration time would be in this case a function of $M$. In Lennard-Jones liquids, this threshold $E_{th}$ has been found to be proportional to the number $N$ of degrees of freedom but, as argued above, $N$ should be actually bounded by $\xi(T)^d$, so that $E_{th}$ is expected to remain finite even for macroscopic systems. Conversely, $M$ should also remain finite, but assuming that the smallest energy $E_{min}$ found in the system lies much below $E_{th}$, one can still take the limit $M \to \infty$ keeping $E_{th}$ fixed, so as to describe the dynamics on time scales much smaller than the finite size equilibration time.

\begin{figure}[t]
\centering\includegraphics[width=8.5cm]{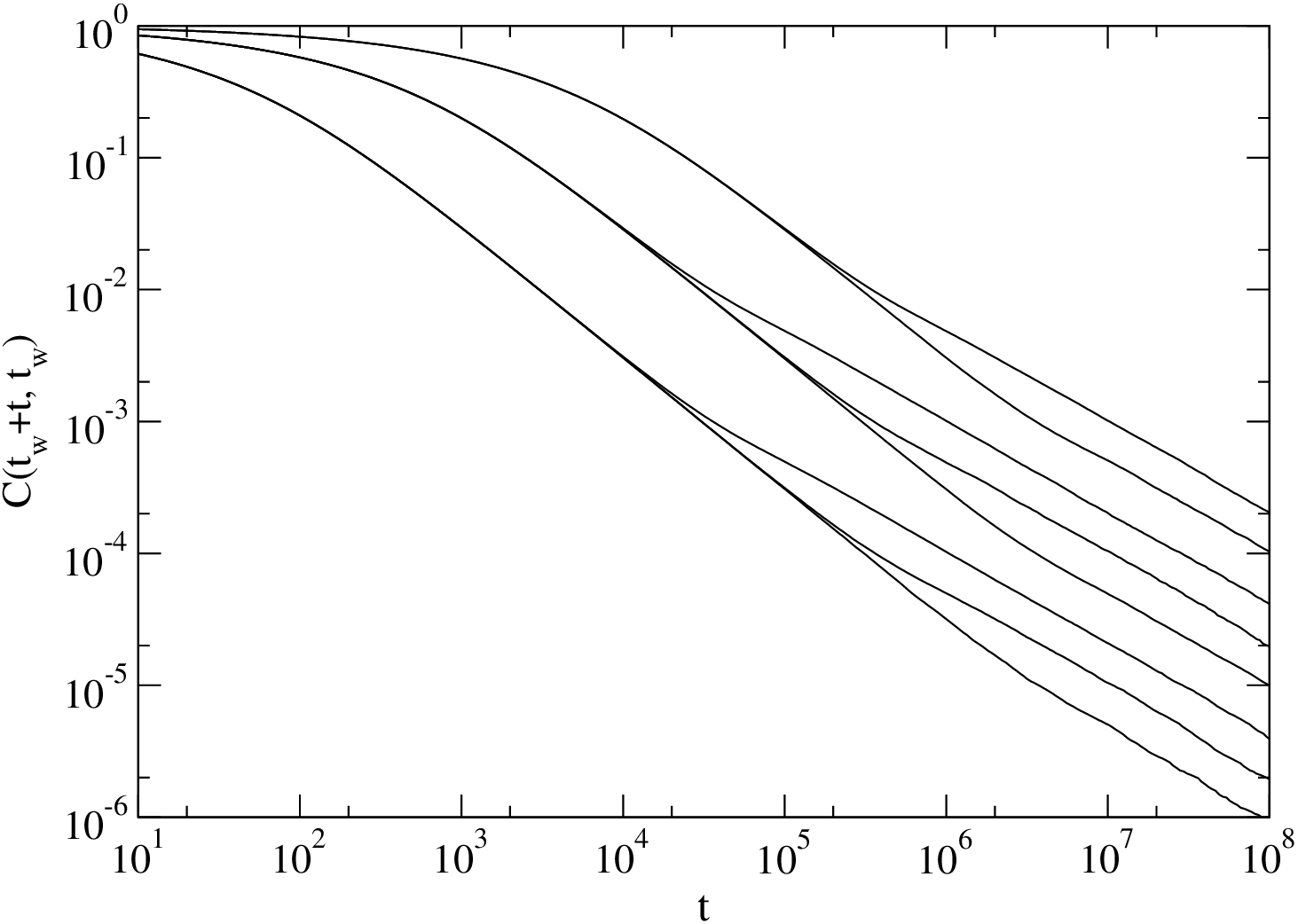}
\caption{\sl Plot of the correlation function $C(t_w,t_w+t)$ of the ThBM, in the aging case ($\mu=0.7$) and in the regime $t \gg t_w$ (long time) and $t_w \ll \tau_{th}$. Both $\tau_{th}$ and $t_w$ are varied: $\tau_{th}=10^5$ ($t_w=10^2$, $10^3$), $10^6$ and $10^7$ ($t_w=10^2$, $10^3$, $10^4$). A crossover appears between an entropic regime, with exponent $-1$, and an activated one, with exponent $-\mu$.
}
\label{fig-thr}
\end{figure}

Fig.~\ref{fig-thr} displays the results of numerical simulations of the correlation function in this generalized model, denoted hereafter as ThBM (`Threshold Barrat-M\'ezard') model, for several values of $t_w$ and $E_{th}$, focusing on the long time behaviour ($t \gg t_w$). The simulations show the onset of a crossover between two different power law behaviours, the first one with an exponent $-1$ corresponding to the usual long time behaviour of the BM model (entropic dynamics), and the other one with an exponent $-\mu$ typical of the long time tail of the correlation in the trap model (thermally activated dynamics).

Let us now present a scaling argument in order to gain a better understanding of this crossover. We first focus on the most interesting case for which the typical energies reached at $t_w$, of the order of $-T_g\, \ln t_w$, remain far enough above $E_{th}$.

\begin{figure}[t]
\centering\includegraphics[width=8.5cm]{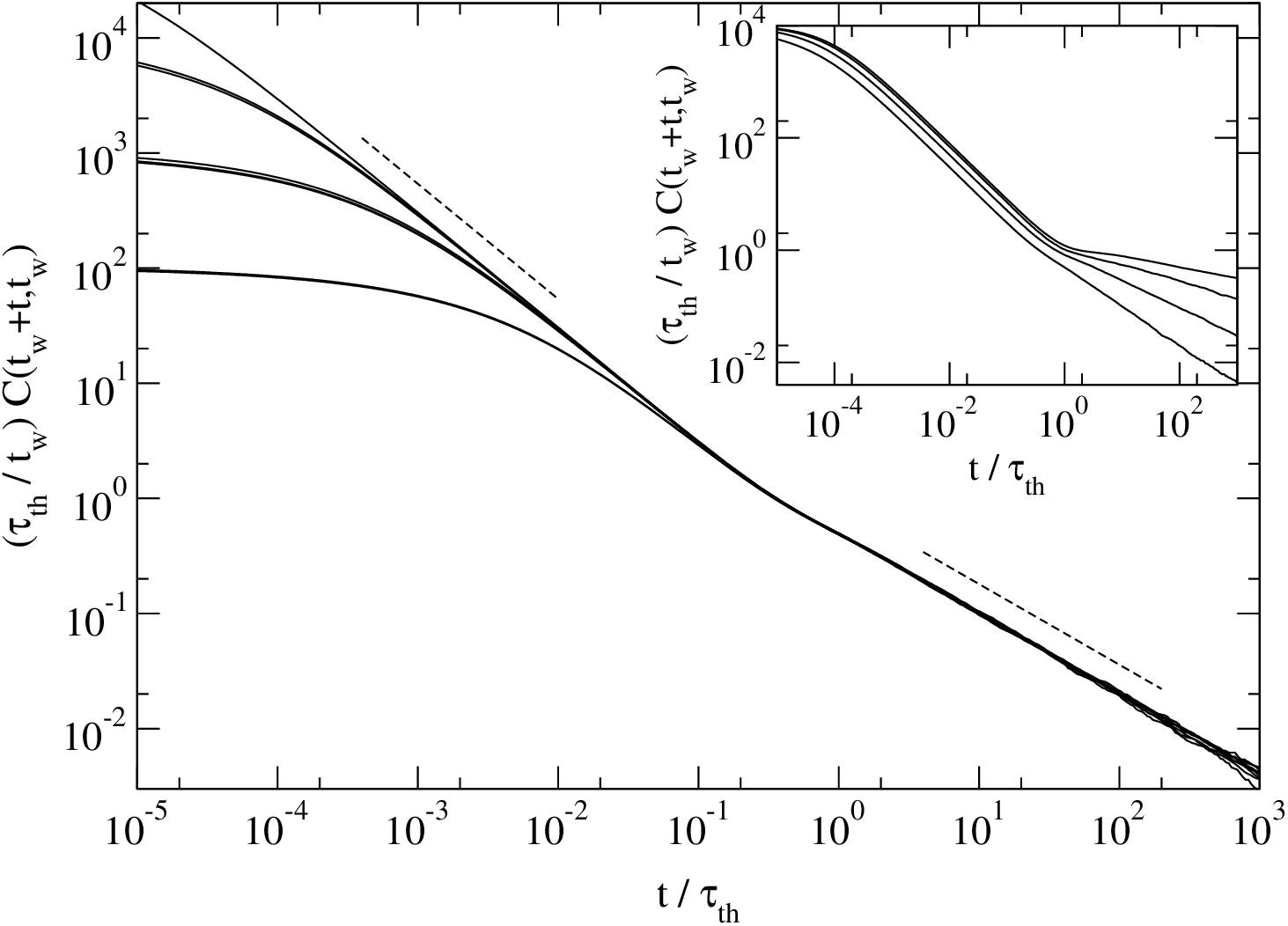}
\caption{\sl Same curves as on Fig.~\ref{fig-thr}, rescaled in their late part in order to evidence the crossover scaling, by plotting $(\tau_{th}/t_w) \, C(t_w,t_w+t)$ versus $t/\tau_{th}$ (see text). Dashed lines indicate slopes $-1$ and $-\mu$ respectively. Inset: rescaled correlation function for $\mu=0.2$, $0.3$, $0.5$ and $0.7$ (from top to bottom), $t_w=10^3$ and $\tau_{th}=10^7$, showing the strong temperature dependence for $t \gg \tau_{th}$, whereas the slope for $t_w \ll t \ll \tau_{th}$ is independent of $\mu$.
}
\label{fig-thr-sc}
\end{figure}

\begin{figure}[t]
\centering\includegraphics[width=8.5cm]{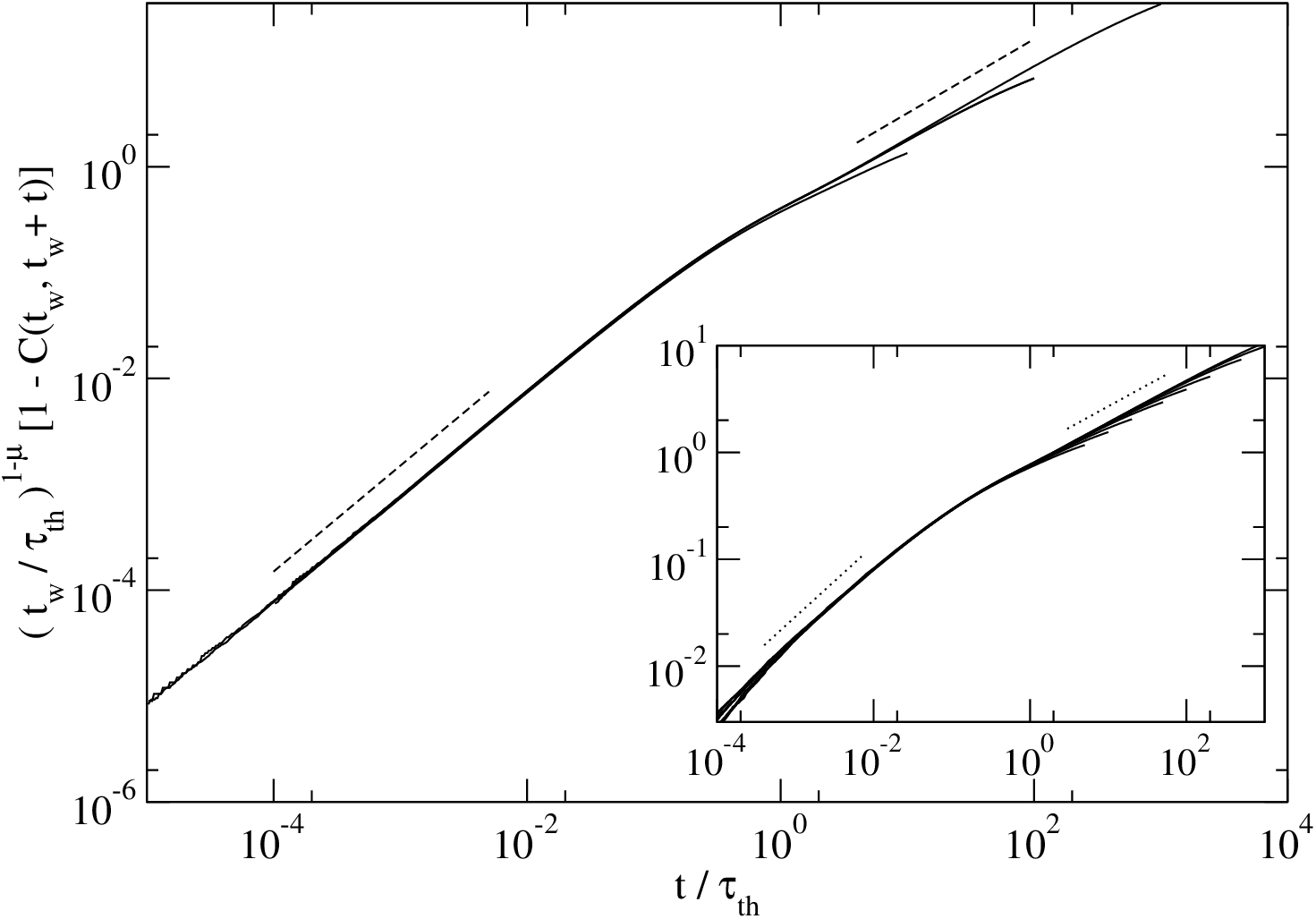}
\caption{\sl Short time behaviour of $C(t_w,t_w+t)$ for $\mu=0.3$ and $t_w \gg \tau_{th}$, plotted as $(t_w/\tau_{th})^{1-\mu} \,[1-C(t_w,t_w+t)]$ versus $t/\tau_{th}$ [see Eq.~(\ref{eqscthsh})] so as to evidence the crossover from a linear (entropic) behaviour for $t \ll \tau_{th}$ to an activated one, with exponent $1-\mu$, for $\tau_{th} \ll t \ll t_w$; $\tau_{th}=10^3$ ($t_w=10^5$, $10^6$) and $10^4$ ($t_w=10^6$, $10^7$). Inset: same plot with $\mu=0.6$, such that the entropic regime $1 \ll t \ll \tau_{th}$ corresponds to a power law with exponent $(1-\mu)/\mu$; $\tau_{th}=5.10^3$, $10^4$, $2.10^4$ and $t_w=10^5$, $10^6$, $10^7$.
}
\label{fig-thr-short3}
\end{figure}

We assume, and we have checked numerically, that the tail ($E$ much below $-T_g\, \ln t_w$) of the dynamical energy distribution $P(E,t_w)$ keeps the usual form $P(E,t_w) \propto e^{E/T_g}$ found both in the BM and in the trap model, for $E > E_{th}$ as well as for $E < E_{th}$. The natural time versus energy relation in the BM model allows to define a time scale $\tau_{th}$ associated to $E_{th}$ through $\tau_{th} = \exp(|E_{th}|/T_g)$ -- we neglect here the prefactor appearing in Eq.~(\ref{eq-tauE}). One thus expects the following asymptotic behaviour for the dynamical distribution $p(\tau,t_w)$ of sojourn times:
\bea
p(\tau,t_w) &\sim& \frac{1}{t_w}\, \left(\frac{t_w}{\tau}\right)^2 \qquad \quad \tau \ll \tau_{th} \\
p(\tau,t_w) &\sim& \frac{A}{t_w}\, \left(\frac{t_w}{\tau}\right)^{1+\mu} \qquad \tau \gg \tau_{th}
\eea
these expressions corresponding to the large $\tau$ tail of the dynamical distribution in the usual BM model and in the trap model respectively. The rescaling factor $A$ is determined by matching the two asymptotic expressions for $\tau=\tau_{th}$, which yields $A = (t_w/\tau_{th})^{1-\mu}$. Computing the correlation for $t \ll \tau_{th}$, one recovers as previously $C(t_w,t_w+t) \sim t_w/t$, whereas for $t \gg \tau_{th}$, one now has:
\be
C(t_w,t_w+t) \sim \frac{t_w}{\tau_{th}}\, \left(\frac{\tau_{th}}{t}\right)^{\mu}
\ee
Summarizing the above results, one can introduce a scaling function $f(.)$ such that $C(t_w,t_w+t)$ reads:
\be \label{EqCsc-thr}
C(t_w,t_w+t) = \frac{t_w}{\tau_{th}} \, f\left(\frac{t}{\tau_{th}}\right)
\ee
with the following asymptotic behaviour for $f(z)$:
\be
f(z) \sim \frac{1}{z} \quad (z \ll 1), \quad f(z) \sim \frac{1}{z^{\mu}} \quad (z \gg 1)
\ee
In order to test this prediction, we have plotted on Fig.~\ref{fig-thr-sc} the previous curves using the rescaled variables $(\tau_{th}/t_w)\, C(t_w,t_w+t)$ and $t/\tau_{th}$. The resulting collapse of the late part of the curves (i.e. for $t \gg t_w$) is very good, confirming the validity of the above scaling analysis. Note that the whole curves cannot be rescaled in a simple way since the early part of the correlation still scales in $t/t_w$ as usual (rescaling not shown), so that two different scaling regimes $t \sim t_w$ and $t \sim \tau_{th}$ have to be considered separately. Fig.~\ref{fig-thr-sc} shows that in the limit of a large separation of scales between $t_w$ and $\tau_{th}$, the plot would reduce to two straight lines with different slopes $-1$ and $-\mu$. The temperature dependence is clearly evidenced in the inset of Fig.~\ref{fig-thr-sc}, showing as expected that the slope is independent of $\mu$ for $t_w \ll t \ll \tau_{th}$ (and equal to $-1$), whereas it is strongly temperature dependent (and equal to $-\mu$) for $t \gg \tau_{th}$.

In the opposite regime $t_w \gg \tau_{th}$, the system is trapped into deep minima so that the behaviour becomes similar to that of the trap model, up to an energy shift. However, the particle necessarily visits states with energy $E > E_{th}$ in order to escape from minima, and then relaxes towards another minima. So a crossover between entropic and activated dynamics is also expected in this case for $t \sim \tau_{th}$, which now falls in the short time regime ($1 \ll t \ll t_w$). In order to analyse this crossover, one can derive a scaling relation along the same lines as for Eq.~(\ref{EqCsc-thr}), which now reads:
\be \label{eqscthsh}
1-C(t_w,t_w+t) = \left( \frac{\tau_{th}}{t_w} \right)^{1-\mu} \, g\left( \frac{t}{\tau_{th}} \right)
\ee
where the scaling function $g(z)$ has the following asymptotic behaviour:
\be
g(z) \sim z^{\gamma} \quad (z \ll 1), \qquad g(z) \sim z^{1-\mu} \quad (z \gg 1)
\ee
with $\gamma = \min[1,(1-\mu)/\mu]$, as for the usual Barrat-M\'ezard model. Fig.~\ref{fig-thr-short3} displays the numerical results for different $t_w$ and $\tau_{th}$, with $\mu=0.3$ and $0.6$, and shows a good collapse of the data, which clearly supports the existence of a crossover for $t \sim \tau_{th}$.

In summary, a non fully connected phase space structure generates a large number of local minima, so that the dynamical crossover time scale associated to the end of the entropic regime is strongly reduced with respect to the fully connected case, suggesting that this crossover may be relevant in realistic glasses on experimental time scales, even below the glass transition temperature.

\section{Conclusion}

In this paper, we have shown that a crossover from entropic to activated behaviour arises naturally in glassy models once a finite coherence length is taken into account, allowing to conceptually decompose a large system into a collection of independent small subsystems \cite{JP-Houches,Biroli}. In this framework, one is then naturally led to consider {\it finite size} phase space models to describe the local dynamics within an `elementary correlated cell'.

Interestingly, calculations in the trap model show that the finite size $M$ replaces in a rough sense the time cut-off $t_w$ appearing in the aging regime, without affecting the asymptotic power law short and late time behaviour of the correlation. On the contrary, in the BM model, the finite size equilibrium behaviour differs significantly from the aging one: the short time singularity remains the same, but the long time one is no longer governed by an entropic mechanism, but corresponds to a thermally activated dynamics since escaping from the deepest states is only possible through jumps to higher states, if the number of states is finite. Generalizing this analysis to the aging regime of finite size systems, where both $t_w$ and $M$ come into play ($t_w \ll M$), the previous crossover still exists for $t \sim M$ as in equilibrium, and can even be found earlier if a finite connectivity is taken into account. Although such a crossover has already been observed in numerical simulations \cite{Angelani,Reichman}, a quantitative analysis of the associated time scales would be very valuable, in order to see how it varies with temperature and with the total size of the system. In particular, if a decomposition into small correlated cells holds, one expects the crossover time scale to saturate to a finite value for large system sizes.

Following the phase space decomposition into saddles and minima, we have then studied a generalization of the BM model in which the crossover time scale is not induced by the finite size of the system, but instead by a threshold energy $E_{th}$ which separates saddles for $E > E_{th}$ from minima for $E < E_{th}$. Assuming that the equilibration time is large enough, we find that the correlation function keeps aging even beyond the crossover time scale $\tau_{th}$, but with a behaviour characteristic of activation dynamics. Interestingly, the transition from one type of aging dynamics to the other leads to an unusual scaling, due to the fact that two time scales $t_w$ and $\tau_{th}$ are now involved instead of one. Moreover, the presence of a lot of local minima in phase space leads to a crossover time scale much smaller than the equilibration time, contrary to what happens in the fully connected case.

\section*{Acknowledgments}

The author gratefully acknowledges J.-P.~Bouchaud for many important and inspiring discussions, to which the present work owes a lot, as well as for a critical reading of the manuscript.

\end{document}